# Compounding Fields and Their Quantum Equations in the Trigintaduonion Space

Zihua Weng

*School of Physics and Mechanical & Electrical Engineering, P. O. Box 310, Xiamen University, Xiamen 361005, China*

**Abstract**

The 32-dimensional compounding fields and their quantum interplays in the trigintaduonion space can be presented by analogy with octonion and sedenion electromagnetic, gravitational, strong and weak interactions. In the trigintaduonion fields which are associated with the electromagnetic, gravitational, strong and weak interactions, the study deduces some conclusions of field source particles (quarks and leptons) and intermediate particles which are consistent with current some sorts of interaction theories. In the trigintaduonion fields which are associated with the hyper-strong and strong-weak fields, the paper draws some predicts and conclusions of the field source particles (sub-quarks) and intermediate particles. The research results show that there may exist some new particles in the nature.

## 1. Introduction

Nowadays, there still exist some movement phenomena which can't be explained by current sub-quark field theories. Therefore, some scientists bring forward some new field theories to explain the strange phenomena in the strong interaction and weak interaction etc.

A new insight on the problem of the sub-quark movement and their interactions can be given by the concept of trigintaduonion space. According to previous research results and the 'SpaceTime Equality Postulation' [1-5], the eight sorts of interactions in the paper can all be described by quaternion spacetimes. Based on the conception of space verticality etc., these eight types of quaternion spacetimes can be united into the 32-dimensional trigintaduonion space. In the trigintaduonion space, the characteristics of eight sorts of interactions can be described by single trigintaduonion space uniformly.

By analogy with the octonionic and sedenion fields, four sorts of trigintaduonion fields which consist of octonionic fields H-S, S-W and E-G etc., can be obtained in the paper. The paper describes the trigintaduonion fields and their quantum theory, and deduces some predicts and new conclusions which are consistent with the current sub-quark theories etc.

———————

*E-mail Addresses*: xmuwzh@hotmail.com, xmuwzh@xmu.edu.cn

## 2. Compounding fields in trigintaduonion spaces

Through the analysis of the different fields in the octonionic and sedenion spaces, we find that each interaction possesses its own spacetime, field and operator in accordance with the 'SpaceTime Equality Postulation'. In the sedenion spaces, sixteen sorts of sedenion fields can be tabulated in Table 1, including their operators, spaces and fields.

Table 1. The compounding fields and operators in the different sedenion spaces

| | | | | |
|---|---|---|---|---|
| operator | $\mathcal{X}_{H-S} / k^X_{H-S} + \diamondsuit$ | $\diamondsuit + \mathcal{A}_{S-W} / k^A_{S-W}$ $+ \mathcal{A}_{H-S} / k^A_{H-S}$ | $\diamondsuit + \mathcal{B}_{E-G} / k^B_{E-G}$ $+ \mathcal{B}_{H-S} / k^B_{H-S}$ | $\diamondsuit + \mathcal{S}_{H-W} / k^S_{H-W}$ $+ \mathcal{S}_{H-S} / k^S_{H-S}$ |
| space | octonion space H-S | sedenion space SW-HS | sedenion space EG-HS | sedenion space HW-HS |
| field | H-S | SW-HS | EG-HS | HW-HS |
| | | | | |
| operator | $\diamondsuit + \mathcal{X}_{H-S} / k^X_{H-S}$ $+ \mathcal{X}_{S-W} / k^X_{S-W}$ | $\mathcal{A}_{S-W} / k^A_{S-W} + \diamondsuit$ | $\diamondsuit + \mathcal{B}_{E-G} / k^B_{E-G}$ $+ \mathcal{B}_{S-W} / k^B_{S-W}$ | $\diamondsuit + \mathcal{S}_{H-W} / k^S_{H-W}$ $+ \mathcal{S}_{S-W} / k^S_{S-W}$ |
| space | sedenion space HS-SW | octonion space S-W | sedenion space EG-SW | sedenion space HW-SW |
| field | HS-SW | S-W | EG-SW | HW-SW |
| | | | | |
| operator | $\diamondsuit + \mathcal{X}_{H-S} / k^X_{H-S}$ $+ \mathcal{X}_{E-G} / k^X_{E-G}$ | $\diamondsuit + \mathcal{A}_{S-W} / k^A_{S-W}$ $+ \mathcal{A}_{E-G} / k^A_{E-G}$ | $\mathcal{B}_{E-G} / k^B_{E-G} + \diamondsuit$ | $\diamondsuit + \mathcal{S}_{H-W} / k^S_{H-W}$ $+ \mathcal{S}_{E-G} / k^S_{E-G}$ |
| space | sedenion space HS-EG | sedenion space SW-EG | octonion space E-G | sedenion space HW-EG |
| field | HS-EG | SW-EG | E-G | HW-EG |
| | | | | |
| operator | $\diamondsuit + \mathcal{X}_{H-S} / k^X_{H-S}$ $+ \mathcal{X}_{H-W} / k^X_{H-W}$ | $\diamondsuit + \mathcal{A}_{S-W} / k^A_{S-W}$ $+ \mathcal{A}_{H-W} / k^A_{H-W}$ | $\diamondsuit + \mathcal{B}_{E-G} / k^B_{E-G}$ $+ \mathcal{B}_{H-W} / k^B_{H-W}$ | $\mathcal{S}_{H-W} / k^S_{H-W} + \diamondsuit$ |
| space | sedenion space HS-HW | sedenion space SW-HW | sedenion space EG-HW | octonion space H-W |
| field | HS-HW | SW-HW | EG-HW | H-W |

In the Cayley-Dickson algebra, there exists the Cayley-Dickson construction [6]. This is the process based on which the 2n-dimensional hypercomplex number is constructed from a pair of (2n-1)-dimensional hypercomplex numbers, where n is a positive integer. This is accomplished by defining the multiplication rule for the two 2n-dimensional hypercomplex numbers in terms of the four (2n-1)-dimensional hypercomplex numbers. The 2-dimensional complex numbers (n = 1), 4-dimensional quaternions (n = 2), 8-dimensional octonions (n = 3), 16-dimensional sedenions (n = 4), 32-dimensional trigintaduonions (n = 5), etc., can all be constructed from real numbers by the iterations of this process [7]. At each iteration some new basal elements, $\vec{e}_k$, are introduced with the property, $\vec{e}_k^2 = -1$.

We define the product and conjugate on the trigintaduonions, (u, v) and (x, y), in terms of the sedenions, u, v, x and y, as follows:

$$(u, v)(x, y) = (ux - y^* v, yu + v x^*), \quad (u, v)^* = (u^*, -v)$$

where, the mark (*) denotes the conjugate.

In the trigintaduonion space, there exist different constructions of fields in the terms of different operators. By analogy with the cases in the different octonionic spaces and sedenion spaces, the operators and fields in the different trigintaduonion spaces can be written in Table 2. There exist four sorts of compounding fields in the trigintaduonion spaces.

Table 2. The compounding fields and operators in the trigintaduonion space

| | $\Diamond + \mathcal{X}_{H-S} / k^X_{H-S}$ $+ \mathcal{X}_{S-W} / k^X_{S-W}$ $+ \mathcal{X}_{E-G} / k^X_{E-G}$ $+ \mathcal{X}_{H-W} / k^X_{H-W}$ | $\Diamond + \mathcal{A}_{H-S} / k^A_{H-S}$ $+ \mathcal{A}_{S-W} / k^A_{S-W}$ $+ \mathcal{A}_{E-G} / k^A_{E-G}$ $+ \mathcal{A}_{H-W} / k^A_{H-W}$ | $\Diamond + \mathcal{B}_{H-S} / k^B_{H-S}$ $+ \mathcal{B}_{S-W} / k^B_{S-W}$ $+ \mathcal{B}_{E-G} / k^B_{E-G}$ $+ \mathcal{B}_{H-W} / k^B_{H-W}$ | $\Diamond + \mathcal{S}_{H-S} / k^S_{H-S}$ $+ \mathcal{S}_{S-W} / k^S_{S-W}$ $+ \mathcal{S}_{E-G} / k^S_{E-G}$ $+ \mathcal{S}_{H-W} / k^S_{H-W}$ |
|---|---|---|---|---|
| operator | | | | |
| space | T-X | T-A | T-B | T-S |
| field | T-X | T-A | T-B | T-S |

## 3. Compounding field in trigintaduonion space T-X

It is believed that hyper-strong field, strong-weak field, electromagnetic-gravitational field and hyper-weak field are unified, equal and interconnected. By means of the conception of the space expansion etc., four types of octonionic spaces can be combined into a trigintaduonion space T-X. In trigintaduonion space, some properties of eight sorts of interactions including strong, weak, electromagnetic and gravitational interactions etc. can be described uniformly.

In the trigintaduonion space T-X, the displacement $r$ should be extended to the new displacement $\mathcal{R} = (r + k_{rx} \mathcal{X})$ and be consistent with the definition of momentum $\mathcal{M}$.

In the octonionic space H-S, the base $\mathcal{E}_{H-S}$ can be written as

$$\mathcal{E}_{H-S} = (\mathbf{1}, \vec{e}_1, \vec{e}_2, \vec{e}_3, \vec{e}_4, \vec{e}_5, \vec{e}_6, \vec{e}_7) \qquad (1)$$

The displacement $\mathcal{R}_{H-S} = (R_0, R_1, R_2, R_3, R_4, R_5, R_6, R_7)$ in the octonionic space H-S is consist of the displacement $r_{H-S} = (r_0, r_1, r_2, r_3, r_4, r_5, r_6, r_7)$ and physical quantity $\mathcal{X}_{H-S} = (x_0, x_1, x_2, x_3, x_4, x_5, x_6, x_7)$.

$$\mathcal{R}_{H-S} = r_{H-S} + k^{H-S}_{rx} \mathcal{X}_{H-S}$$
$$= R_0 + \vec{e}_1 R_1 + \vec{e}_2 R_2 + \vec{e}_3 R_3 + \vec{e}_4 R_4 + \vec{e}_5 R_5 + \vec{e}_6 R_6 + \vec{e}_7 R_7 \qquad (2)$$

where, $R_j = r_j + k^{H-S}_{rx} x_j$ ; $j = 0, 1, 2, 3, 4, 5, 6, 7$. $r_0 = c_{H-S} t_{H-S}$, $r_4 = c_{H-S} T_{H-S}$. $c_{H-S}$ is the speed of intermediate particle in the hyper-strong field, $t_{H-S}$ and $T_{H-S}$ denote the time.

The octonionic differential operator $\Diamond_{T-X1}$ and its conjugate operator are defined as

$$\Diamond_{T-X1} = \partial_0 + \vec{e}_1 \partial_1 + \vec{e}_2 \partial_2 + \vec{e}_3 \partial_3 + \vec{e}_4 \partial_4 + \vec{e}_5 \partial_5 + \vec{e}_6 \partial_6 + \vec{e}_7 \partial_7 \qquad (3)$$
$$\Diamond^*_{T-X1} = \partial_0 - \vec{e}_1 \partial_1 - \vec{e}_2 \partial_2 - \vec{e}_3 \partial_3 - \vec{e}_4 \partial_4 - \vec{e}_5 \partial_5 - \vec{e}_6 \partial_6 - \vec{e}_7 \partial_7 \qquad (4)$$

where, $\partial_j = \partial / \partial R_j$. The mark (*) denotes the octonionic conjugate.

In the octonionic space S-W, the base $\mathcal{E}_{S-W}$ can be written as

$$\mathcal{E}_{S-W} = (\vec{e}_8, \vec{e}_9, \vec{e}_{10}, \vec{e}_{11}, \vec{e}_{12}, \vec{e}_{13}, \vec{e}_{14}, \vec{e}_{15}) \qquad (5)$$

The displacement $\mathcal{R}_{S-W} = (R_8, R_9, R_{10}, R_{11}, R_{12}, R_{13}, R_{14}, R_{15})$ in the octonionic space S-W is consist of the displacement $r_{S-W} = (r_8, r_9, r_{10}, r_{11}, r_{12}, r_{13}, r_{14}, r_{15})$ and the physical quantity $\mathcal{X}_{S\text{-}W} = (x_8, x_9, x_{10}, x_{11}, x_{12}, x_{13}, x_{14}, x_{15})$.

$$\begin{aligned}\mathcal{R}_{S\text{-}W} &= \mathcal{r}_{S\text{-}W} + k^{S\text{-}W}_{rx}\mathcal{X}_{S\text{-}W}\\ &= \vec{e}_8 R_8 + \vec{e}_9 R_9 + \vec{e}_{10} R_{10} + \vec{e}_{11} R_{11}\\ &\quad + \vec{e}_{12} R_{12} + \vec{e}_{13} R_{13} + \vec{e}_{14} R_{14} + \vec{e}_{15} R_{15}\end{aligned} \quad (6)$$

where, $R_j = r_j + k^{S\text{-}W}_{rx} x_j$; j = 8, 9, 10, 11, 12, 13, 14, 15. $r_8 = c_{S\text{-}W} t_{S\text{-}W}$, $r_{12} = c_{S\text{-}W} T_{S\text{-}W}$. $c_{S\text{-}W}$ is the speed of intermediate particle in strong-weak field, $t_{S\text{-}W}$ and $T_{S\text{-}W}$ denote the time.

The octonionic differential operator $\diamondsuit_{T\text{-}X2}$ and its conjugate operator are defined as

$$\begin{aligned}\diamondsuit_{T\text{-}X2} &= \vec{e}_8 \partial_8 + \vec{e}_9 \partial_9 + \vec{e}_{10} \partial_{10} + \vec{e}_{11} \partial_{11}\\ &\quad + \vec{e}_{12} \partial_{12} + \vec{e}_{13} \partial_{13} + \vec{e}_{14} \partial_{14} + \vec{e}_{15} \partial_{15}\end{aligned} \quad (7)$$

$$\begin{aligned}\diamondsuit^*_{T\text{-}X2} &= -\vec{e}_8 \partial_8 - \vec{e}_9 \partial_9 - \vec{e}_{10} \partial_{10} - \vec{e}_{11} \partial_{11}\\ &\quad - \vec{e}_{12} \partial_{12} - \vec{e}_{13} \partial_{13} - \vec{e}_{14} \partial_{14} - \vec{e}_{15} \partial_{15}\end{aligned} \quad (8)$$

where, $\partial_j = \partial/\partial R_j$.

In the octonionic space E-G, the base $\mathcal{E}_{E\text{-}G}$ can be written as

$$\mathcal{E}_{E\text{-}G} = (\vec{e}_{16}, \vec{e}_{17}, \vec{e}_{18}, \vec{e}_{19}, \vec{e}_{20}, \vec{e}_{21}, \vec{e}_{22}, \vec{e}_{23}) \quad (9)$$

The displacement $\mathcal{R}_{E\text{-}G} = (R_{16}, R_{17}, R_{18}, R_{19}, R_{20}, R_{21}, R_{22}, R_{23})$ in the octonionic space E-G is consist of the displacement $\mathcal{r}_{E\text{-}G} = (r_{16}, r_{17}, r_{18}, r_{19}, r_{20}, r_{21}, r_{22}, r_{23})$ and the physical quantity $\mathcal{X}_{E\text{-}G} = (x_{16}, x_{17}, x_{18}, x_{19}, x_{20}, x_{21}, x_{22}, x_{23})$.

$$\begin{aligned}\mathcal{R}_{E\text{-}G} &= \mathcal{r}_{E\text{-}G} + k^{E\text{-}G}_{rx}\mathcal{X}_{E\text{-}G}\\ &= \vec{e}_{16} R_{16} + \vec{e}_{17} R_{17} + \vec{e}_{18} R_{18} + \vec{e}_{19} R_{19}\\ &\quad + \vec{e}_{20} R_{20} + \vec{e}_{21} R_{21} + \vec{e}_{22} R_{22} + \vec{e}_{23} R_{23}\end{aligned} \quad (10)$$

where, $R_j = r_j + k^{E\text{-}G}_{rx} x_j$; j = 16, 17, 18, 19, 20, 21, 22, 23. $r_{16} = c_{E\text{-}G} t_{E\text{-}G}$, $r_{20} = c_{E\text{-}G} T_{E\text{-}G}$. $c_{E\text{-}G}$ is the speed of intermediate particle in electromagnetic-gravitational field, $t_{E\text{-}G}$ and $T_{E\text{-}G}$ denote the time.

The octonionic differential operator $\diamondsuit_{T\text{-}X3}$ and its conjugate operator are defined as

$$\begin{aligned}\diamondsuit_{T\text{-}X3} &= \vec{e}_{16} \partial_{16} + \vec{e}_{17} \partial_{17} + \vec{e}_{18} \partial_{18} + \vec{e}_{19} \partial_{19}\\ &\quad + \vec{e}_{20} \partial_{20} + \vec{e}_{21} \partial_{21} + \vec{e}_{22} \partial_{22} + \vec{e}_{23} \partial_{23}\end{aligned} \quad (11)$$

$$\begin{aligned}\diamondsuit^*_{T\text{-}X3} &= -\vec{e}_{16} \partial_{16} - \vec{e}_{17} \partial_{17} - \vec{e}_{18} \partial_{18} - \vec{e}_{19} \partial_{19}\\ &\quad - \vec{e}_{20} \partial_{20} - \vec{e}_{21} \partial_{21} - \vec{e}_{22} \partial_{22} - \vec{e}_{23} \partial_{23}\end{aligned} \quad (12)$$

where, $\partial_j = \partial/\partial R_j$.

In the octonionic space H-W, the base $\mathcal{E}_{H\text{-}W}$ can be written as

$$\mathcal{E}_{H\text{-}W} = (\vec{e}_{24}, \vec{e}_{25}, \vec{e}_{26}, \vec{e}_{27}, \vec{e}_{28}, \vec{e}_{29}, \vec{e}_{30}, \vec{e}_{31}) \quad (13)$$

The displacement $\mathcal{R}_{H\text{-}W} = (R_{24}, R_{25}, R_{26}, R_{27}, R_{28}, R_{29}, R_{30}, R_{31})$ in octonionic space H-W is consist of the displacement $\mathcal{r}_{H\text{-}W} = (r_{24}, r_{25}, r_{26}, r_{27}, r_{28}, r_{29}, r_{30}, r_{31})$ and the physical quantity $\mathcal{X}_{H\text{-}W} = (x_{24}, x_{25}, x_{26}, x_{27}, x_{28}, x_{29}, x_{30}, x_{31})$.

$$\begin{aligned}\mathcal{R}_{H\text{-}W} &= \mathcal{r}_{H\text{-}W} + k^{H\text{-}W}_{rx}\mathcal{X}_{H\text{-}W}\\ &= \vec{e}_{24} R_{24} + \vec{e}_{25} R_{25} + \vec{e}_{26} R_{26} + \vec{e}_{27} R_{27}\\ &\quad + \vec{e}_{28} R_{28} + \vec{e}_{29} R_{29} + \vec{e}_{30} R_{30} + \vec{e}_{31} R_{31}\end{aligned} \quad (14)$$

where, $R_j = r_j + k^{H\text{-}W}_{rx} x_j$; j = 24, 25, 26, 27, 28, 29, 30, 31. $r_{24} = c_{H\text{-}W} t_{H\text{-}W}$, $r_{28} = c_{H\text{-}W} T_{H\text{-}W}$. $c_{H\text{-}W}$ is the speed of intermediate particle in hyper-weak field, $t_{H\text{-}W}$ and $T_{H\text{-}W}$ denote the time.

The octonionic differential operator $\diamondsuit_{T\text{-}X4}$ and its conjugate operator are defined as,

$$\begin{aligned}\diamondsuit_{T\text{-}X4} &= \vec{e}_{24} \partial_{24} + \vec{e}_{25} \partial_{25} + \vec{e}_{26} \partial_{26} + \vec{e}_{27} \partial_{27}\\ &\quad + \vec{e}_{28} \partial_{28} + \vec{e}_{29} \partial_{29} + \vec{e}_{30} \partial_{30} + \vec{e}_{31} \partial_{31}\end{aligned} \quad (15)$$

$$\diamondsuit^*_{T\text{-}X4} = -\vec{e}_{24} \partial_{24} - \vec{e}_{25} \partial_{25} - \vec{e}_{26} \partial_{26} - \vec{e}_{27} \partial_{27}$$

$$-\vec{e}_{28}\partial_{28} - \vec{e}_{29}\partial_{29} - \vec{e}_{30}\partial_{30} - \vec{e}_{31}\partial_{31} \tag{16}$$

where, $\partial_j = \partial/\partial R_j$.

In the trigintaduonion space T-X, the base $\mathcal{E}_{T-X}$ can be written as

$$\begin{aligned}\mathcal{E}_{T-X} &= \mathcal{E}_{T-X1} + \mathcal{E}_{T-X2} + \mathcal{E}_{T-X3} + \mathcal{E}_{T-X4}\\ &= (\mathbf{1}, \vec{e}_1, \vec{e}_2, \vec{e}_3, \vec{e}_4, \vec{e}_5, \vec{e}_6, \vec{e}_7, \vec{e}_8, \vec{e}_9, \vec{e}_{10},\\ &\quad \vec{e}_{11}, \vec{e}_{12}, \vec{e}_{13}, \vec{e}_{14}, \vec{e}_{15}, \vec{e}_{16}, \vec{e}_{17}, \vec{e}_{18}, \vec{e}_{19}, \vec{e}_{20},\\ &\quad \vec{e}_{21}, \vec{e}_{22}, \vec{e}_{23}, \vec{e}_{24}, \vec{e}_{25}, \vec{e}_{26}, \vec{e}_{27}, \vec{e}_{28}, \vec{e}_{29}, \vec{e}_{30}, \vec{e}_{31})\end{aligned} \tag{17}$$

The displacement $\mathcal{R}_{T-X}$ = ( $R_0$, $R_1$, $R_2$, $R_3$, $R_4$, $R_5$, $R_6$, $R_7$, $R_8$, $R_9$, $R_{10}$, $R_{11}$, $R_{12}$, $R_{13}$, $R_{14}$, $R_{15}$, $R_{16}$, $R_{17}$, $R_{18}$, $R_{19}$, $R_{20}$, $R_{21}$, $R_{22}$, $R_{23}$, $R_{24}$, $R_{25}$, $R_{26}$, $R_{27}$, $R_{28}$, $R_{29}$, $R_{30}$, $R_{31}$ ) in trigintaduonion space T-X is

$$\begin{aligned}\mathcal{R}_{T-X} &= \mathcal{R}_{T-X1} + \mathcal{R}_{T-X2} + \mathcal{R}_{T-X3} + \mathcal{R}_{T-X4}\\ &= R_0 + \vec{e}_1 R_1 + \vec{e}_2 R_2 + \vec{e}_3 R_3 + \vec{e}_4 R_4 + \vec{e}_5 R_5 + \vec{e}_6 R_6\\ &\quad + \vec{e}_7 R_7 + \vec{e}_8 R_8 + \vec{e}_9 R_9 + \vec{e}_{10} R_{10} + \vec{e}_{11} R_{11}\\ &\quad + \vec{e}_{12} R_{12} + \vec{e}_{13} R_{13} + \vec{e}_{14} R_{14} + \vec{e}_{15} R_{15} + \vec{e}_{16} R_{16}\\ &\quad + \vec{e}_{17} R_{17} + \vec{e}_{18} R_{18} + \vec{e}_{19} R_{19} + \vec{e}_{20} R_{20} + \vec{e}_{21} R_{21}\\ &\quad + \vec{e}_{22} R_{22} + \vec{e}_{23} R_{23} + \vec{e}_{24} R_{24} + \vec{e}_{25} R_{25} + \vec{e}_{26} R_{26}\\ &\quad + \vec{e}_{27} R_{27} + \vec{e}_{28} R_{28} + \vec{e}_{29} R_{29} + \vec{e}_{30} R_{30} + \vec{e}_{31} R_{31}\end{aligned} \tag{18}$$

The trigintaduonion differential operator $\Diamond_{T-X}$ and its conjugate operator are defined as

$$\Diamond_{T-X} = \Diamond_{T-X1} + \Diamond_{T-X2} + \Diamond_{T-X3} + \Diamond_{T-X4} \tag{19}$$

$$\Diamond^*_{T-X} = \Diamond^*_{T-X1} + \Diamond^*_{T-X2} + \Diamond^*_{T-X3} + \Diamond^*_{T-X4} \tag{20}$$

In the trigintaduonion space T-X, there exists one kind of field (trigintaduonion field T-X, for short) can be obtained related to the operator ($\mathcal{X}/K + \Diamond$).

In the trigintaduonion field T-X, by analogy with the octonion and sedenion fields, the trigintaduonion differential operator $\Diamond$ needs to be generalized to the operator ($\mathcal{X}_{H-S}/k^X_{H-S} + \mathcal{X}_{S-W}/k^X_{S-W} + \mathcal{X}_{E-G}/k^X_{E-G} + \mathcal{X}_{H-W}/k^X_{H-W} + \Diamond$). This is because the trigintaduonion field T-X includes the hyper-strong, strong-weak, electromagnetic-gravitational and hyper-weak fields.

It can be predicted that the eight sorts of interactions are interconnected each other. The physical features of each subfield in the trigintaduonion field T-X meet the requirements of the equations set in the Table 3.

In the trigintaduonion field T-X, the field potential $\mathcal{A}$ = ($a_0$, $a_1$, $a_2$, $a_3$, $a_4$, $a_5$, $a_6$, $a_7$, $a_8$, $a_9$, $a_{10}$, $a_{11}$, $a_{12}$, $a_{13}$, $a_{14}$, $a_{15}$, $a_{16}$, $a_{17}$, $a_{18}$, $a_{19}$, $a_{20}$, $a_{21}$, $a_{22}$, $a_{23}$, $a_{24}$, $a_{25}$, $a_{26}$, $a_{27}$, $a_{28}$, $a_{29}$, $a_{30}$, $a_{31}$ ) is defined as

$$\begin{aligned}\mathcal{A} &= (\mathcal{X}/K + \Diamond)^* \circ \mathcal{X}\\ &= (\mathcal{X}_{H-S}/k^X_{H-S} + \mathcal{X}_{S-W}/k^X_{S-W} + \mathcal{X}_{E-G}/k^X_{E-G} + \mathcal{X}_{H-W}/k^X_{H-W} + \Diamond)^* \circ \mathcal{X}\\ &= a_0 + a_1 \vec{e}_1 + a_2 \vec{e}_2 + a_3 \vec{e}_3 + a_4 \vec{e}_4 + a_5 \vec{e}_5 + a_6 \vec{e}_6\\ &\quad + a_7 \vec{e}_7 + a_8 \vec{e}_8 + a_9 \vec{e}_9 + a_{10} \vec{e}_{10} + a_{11} \vec{e}_{11}\\ &\quad + a_{12} \vec{e}_{12} + a_{13} \vec{e}_{13} + a_{14} \vec{e}_{14} + a_{15} \vec{e}_{15} + a_{16} \vec{e}_{16}\\ &\quad + a_{17} \vec{e}_{17} + a_{18} \vec{e}_{18} + a_{19} \vec{e}_{19} + a_{20} \vec{e}_{20} + a_{21} \vec{e}_{21}\\ &\quad + a_{22} \vec{e}_{22} + a_{23} \vec{e}_{23} + a_{24} \vec{e}_{24} + a_{25} \vec{e}_{25} + a_{26} \vec{e}_{26}\\ &\quad + a_{27} \vec{e}_{27} + a_{28} \vec{e}_{28} + a_{29} \vec{e}_{29} + a_{30} \vec{e}_{30} + a_{31} \vec{e}_{31}\end{aligned} \tag{21}$$

where, the mark (*) denotes the trigintaduonion conjugate. $k_{rx} \mathcal{X} = k_{rx} \mathcal{X}_{T-X} = k^{H-S}_{rx} \mathcal{X}_{H-S} + k^{S-W}_{rx} \mathcal{X}_{S-W} + k^{E-G}_{rx} \mathcal{X}_{E-G} + k^{H-W}_{rx} \mathcal{X}_{H-W}$. $K = K_{T-X}$, $k^X_{H-S}$, $k^X_{S-W}$, $k^X_{E-G}$, $k^X_{H-W}$, $k^{H-S}_{rx}$, $k^{S-W}_{rx}$, $k^{E-G}_{rx}$ and $k^{H-W}_{rx}$ are coefficients. $\mathcal{X}_{H-S}$ is the physical quantity in the octonionic space H-S;

$\mathcal{X}_{\text{S-W}}$ is the physical quantity in octonionic space S-W; $\mathcal{X}_{\text{E-G}}$ is the physical quantity in the octonionic space E-G; $\mathcal{X}_{\text{H-W}}$ is the physical quantity in the octonionic space H-W.

The field strength $\mathcal{B}$ of the trigintaduonion field T-X can be defined as

$$\mathcal{B} = (\mathcal{X}_{\text{H-S}} / k^X_{\text{H-S}} + \mathcal{X}_{\text{S-W}} / k^X_{\text{S-W}} + \mathcal{X}_{\text{E-G}} / k^X_{\text{E-G}} + \mathcal{X}_{\text{H-W}} / k^X_{\text{H-W}} + \diamond) \circ \mathcal{A} \tag{22}$$

The field source and force of the trigintaduonion field T-X can be defined respectively as

$$\mu S = (\mathcal{X}_{\text{H-S}} / k^X_{\text{H-S}} + \mathcal{X}_{\text{S-W}} / k^X_{\text{S-W}} + \mathcal{X}_{\text{E-G}} / k^X_{\text{E-G}} + \mathcal{X}_{\text{H-W}} / k^X_{\text{H-W}} + \diamond)^* \circ \mathcal{B} \tag{23}$$

$$Z = K (\mathcal{X}_{\text{H-S}} / k^X_{\text{H-S}} + \mathcal{X}_{\text{S-W}} / k^X_{\text{S-W}} + \mathcal{X}_{\text{E-G}} / k^X_{\text{E-G}} + \mathcal{X}_{\text{H-W}} / k^X_{\text{H-W}} + \diamond) \circ S \tag{24}$$

where, the coefficient μ is interaction intensity of the trigintaduonion field T-X.

The angular momentum of trigintaduonion field can be defined as ($k_{rx}$ is the coefficient)

$$\mathcal{M} = S \circ (r + k_{rx} \mathcal{X}) \tag{25}$$

and the energy and power in the trigintaduonion field can be defined respectively as

$$\mathcal{W} = K (\mathcal{X}_{\text{H-S}} / k^X_{\text{H-S}} + \mathcal{X}_{\text{S-W}} / k^X_{\text{S-W}} + \mathcal{X}_{\text{E-G}} / k^X_{\text{E-G}} + \mathcal{X}_{\text{H-W}} / k^X_{\text{H-W}} + \diamond)^* \circ \mathcal{M} \tag{26}$$

$$\mathcal{N} = K (\mathcal{X}_{\text{H-S}} / k^X_{\text{H-S}} + \mathcal{X}_{\text{S-W}} / k^X_{\text{S-W}} + \mathcal{X}_{\text{E-G}} / k^X_{\text{E-G}} + \mathcal{X}_{\text{H-W}} / k^X_{\text{H-W}} + \diamond) \circ \mathcal{W} \tag{27}$$

Table 3.   Equations set of trigintaduonion field T-X

| Spacetime | trigintaduonion space T-X |
|---|---|
| $\mathcal{X}$ physical quantity | $\mathcal{X} = \mathcal{X}_{\text{T-X}}$ |
| Field potential | $\mathcal{A} = (\mathcal{X}_{\text{H-S}} / k^X_{\text{H-S}} + \mathcal{X}_{\text{S-W}} / k^X_{\text{S-W}} + \mathcal{X}_{\text{E-G}} / k^X_{\text{E-G}} + \mathcal{X}_{\text{H-W}} / k^X_{\text{H-W}} + \diamond)^* \circ \mathcal{X}$ |
| Field strength | $\mathcal{B} = (\mathcal{X}_{\text{H-S}} / k^X_{\text{H-S}} + \mathcal{X}_{\text{S-W}} / k^X_{\text{S-W}} + \mathcal{X}_{\text{E-G}} / k^X_{\text{E-G}} + \mathcal{X}_{\text{H-W}} / k^X_{\text{H-W}} + \diamond) \circ \mathcal{A}$ |
| Field source | $\mu S = (\mathcal{X}_{\text{H-S}} / k^X_{\text{H-S}} + \mathcal{X}_{\text{S-W}} / k^X_{\text{S-W}} + \mathcal{X}_{\text{E-G}} / k^X_{\text{E-G}} + \mathcal{X}_{\text{H-W}} / k^X_{\text{H-W}} + \diamond)^* \circ \mathcal{B}$ |
| Force | $Z = K (\mathcal{X}_{\text{H-S}} / k^X_{\text{H-S}} + \mathcal{X}_{\text{S-W}} / k^X_{\text{S-W}} + \mathcal{X}_{\text{E-G}} / k^X_{\text{E-G}} + \mathcal{X}_{\text{H-W}} / k^X_{\text{H-W}} + \diamond) \circ S$ |
| Angular momentum | $\mathcal{M} = S \circ (r + k_{rx} \mathcal{X})$ |
| Energy | $\mathcal{W} = K (\mathcal{X}_{\text{H-S}} / k^X_{\text{H-S}} + \mathcal{X}_{\text{S-W}} / k^X_{\text{S-W}} + \mathcal{X}_{\text{E-G}} / k^X_{\text{E-G}} + \mathcal{X}_{\text{H-W}} / k^X_{\text{H-W}} + \diamond)^* \circ \mathcal{M}$ |
| Power | $\mathcal{N} = K (\mathcal{X}_{\text{H-S}} / k^X_{\text{H-S}} + \mathcal{X}_{\text{S-W}} / k^X_{\text{S-W}} + \mathcal{X}_{\text{E-G}} / k^X_{\text{E-G}} + \mathcal{X}_{\text{H-W}} / k^X_{\text{H-W}} + \diamond) \circ \mathcal{W}$ |

In the trigintaduonion space T-X, the wave functions of the quantum mechanics are the trigintaduonion equations set. The Dirac and Klein-Gordon equations of quantum mechanics are actually the wave equations set which are associated with particle's angular momentum.

In the trigintaduonion field T-X, the Dirac equation and Klein-Gordon equation can be attained respectively from the energy equation (26) and power equation (27) after substituting the operator K ($\mathcal{X}_{\text{H-S}} / k^X_{\text{H-S}} + \mathcal{X}_{\text{S-W}} / k^X_{\text{S-W}} + \mathcal{X}_{\text{E-G}} / k^X_{\text{E-G}} + \mathcal{X}_{\text{H-W}} / k^X_{\text{H-W}} + \diamond$) for the operator ($\mathcal{W}_{\text{H-S}} / k^X_{\text{H-S}} b^X_{\text{H-S}} + \mathcal{W}_{\text{S-W}} / k^X_{\text{S-W}} b^X_{\text{S-W}} + \mathcal{W}_{\text{E-G}} / k^X_{\text{E-G}} b^X_{\text{E-G}} + \mathcal{W}_{\text{H-W}} / k^X_{\text{H-W}} b^X_{\text{H-W}} + \diamond$). The coefficients $b^X_{\text{H-S}}$, $b^X_{\text{S-W}}$, $b^X_{\text{E-G}}$ and $b^X_{\text{H-W}}$ are the Plank-like constant.

The $\mathcal{U}$ equation of the quantum mechanics can be defined as

$$\mathcal{U} = (\mathcal{W}_{\text{H-S}} / k^X_{\text{H-S}} b^X_{\text{H-S}} + \mathcal{W}_{\text{S-W}} / k^X_{\text{S-W}} b^X_{\text{S-W}} + \mathcal{W}_{\text{E-G}} / k^X_{\text{E-G}} b^X_{\text{E-G}} + \mathcal{W}_{\text{H-W}} / k^X_{\text{H-W}} b^X_{\text{H-W}} + \diamond)^* \circ \mathcal{M} \tag{28}$$

The $\mathcal{L}$ equation of the quantum mechanics can be defined as

$$\mathcal{L} = (\mathcal{W}_{\text{H-S}} / k^X_{\text{H-S}} b^X_{\text{H-S}} + \mathcal{W}_{\text{S-W}} / k^X_{\text{S-W}} b^X_{\text{S-W}} + \mathcal{W}_{\text{E-G}} / k^X_{\text{E-G}} b^X_{\text{E-G}} + \mathcal{W}_{\text{H-W}} / k^X_{\text{H-W}} b^X_{\text{H-W}} + \diamond) \circ \mathcal{U} \tag{29}$$

The four sorts of Dirac-like equations can be obtained from the Eqs.(21), (22), (23) and (24) respectively.

The $\mathcal{D}$ equation of quantum mechanics can be defined as

$$\mathcal{D} = (\mathcal{W}_{H-S}/k^X{}_{H-S}b^X{}_{H-S} + \mathcal{W}_{S-W}/k^X{}_{S-W}b^X{}_{S-W}$$
$$+ \mathcal{W}_{E-G}/k^X{}_{E-G}b^X{}_{E-G} + \mathcal{W}_{H-W}/k^X{}_{H-W}b^X{}_{H-W} + \diamondsuit)^* \circ \mathcal{X} \qquad (30)$$

The $\mathcal{G}$ equation of quantum mechanics can be defined as

$$\mathcal{G} = (\mathcal{W}_{H-S}/k^X{}_{H-S}b^X{}_{H-S} + \mathcal{W}_{S-W}/k^X{}_{S-W}b^X{}_{S-W}$$
$$+ \mathcal{W}_{E-G}/k^X{}_{E-G}b^X{}_{E-G} + \mathcal{W}_{H-W}/k^X{}_{H-W}b^X{}_{H-W} + \diamondsuit) \circ \mathcal{D} \qquad (31)$$

The $\mathcal{T}$ equation of quantum mechanics can be defined as

$$\mathcal{T} = (\mathcal{W}_{H-S}/k^X{}_{H-S}b^X{}_{H-S} + \mathcal{W}_{S-W}/k^X{}_{S-W}b^X{}_{S-W}$$
$$+ \mathcal{W}_{E-G}/k^X{}_{E-G}b^X{}_{E-G} + \mathcal{W}_{H-W}/k^X{}_{H-W}b^X{}_{H-W} + \diamondsuit)^* \circ \mathcal{G} \qquad (32)$$

The $O$ equation of quantum mechanics can be defined as

$$O = (\mathcal{W}_{H-S}/k^X{}_{H-S}b^X{}_{H-S} + \mathcal{W}_{S-W}/k^X{}_{S-W}b^X{}_{S-W}$$
$$+ \mathcal{W}_{E-G}/k^X{}_{E-G}b^X{}_{E-G} + \mathcal{W}_{H-W}/k^X{}_{H-W}b^X{}_{H-W} + \diamondsuit) \circ \mathcal{T} \qquad (33)$$

In the trigintaduonion field T-X, the intermediate and field source particles can be obtained. We can find that the intermediate particles and other kinds of new and unknown particles may be existed in the nature.

Table 4.  Quantum equations set of trigintaduonion field T-X

| | |
|---|---|
| Energy quantum | $\mathcal{U} = (\mathcal{W}_{H-S}/k^X{}_{H-S}b^X{}_{H-S} + \mathcal{W}_{S-W}/k^X{}_{S-W}b^X{}_{S-W}$ $+ \mathcal{W}_{E-G}/k^X{}_{E-G}b^X{}_{E-G} + \mathcal{W}_{H-W}/k^X{}_{H-W}b^X{}_{H-W} + \diamondsuit)^* \circ \mathcal{M}$ |
| Power quantum | $\mathcal{L} = (\mathcal{W}_{H-S}/k^X{}_{H-S}b^X{}_{H-S} + \mathcal{W}_{S-W}/k^X{}_{S-W}b^X{}_{S-W}$ $+ \mathcal{W}_{E-G}/k^X{}_{E-G}b^X{}_{E-G} + \mathcal{W}_{H-W}/k^X{}_{H-W}b^X{}_{H-W} + \diamondsuit) \circ \mathcal{U}$ |
| Field potential quantum | $\mathcal{D} = (\mathcal{W}_{H-S}/k^X{}_{H-S}b^X{}_{H-S} + \mathcal{W}_{S-W}/k^X{}_{S-W}b^X{}_{S-W}$ $+ \mathcal{W}_{E-G}/k^X{}_{E-G}b^X{}_{E-G} + \mathcal{W}_{H-W}/k^X{}_{H-W}b^X{}_{H-W} + \diamondsuit)^* \circ \mathcal{X}$ |
| Field strength quantum | $\mathcal{G} = (\mathcal{W}_{H-S}/k^X{}_{H-S}b^X{}_{H-S} + \mathcal{W}_{S-W}/k^X{}_{S-W}b^X{}_{S-W}$ $+ \mathcal{W}_{E-G}/k^X{}_{E-G}b^X{}_{E-G} + \mathcal{W}_{H-W}/k^X{}_{H-W}b^X{}_{H-W} + \diamondsuit) \circ \mathcal{D}$ |
| Field source quantum | $\mathcal{T} = (\mathcal{W}_{H-S}/k^X{}_{H-S}b^X{}_{H-S} + \mathcal{W}_{S-W}/k^X{}_{S-W}b^X{}_{S-W}$ $+ \mathcal{W}_{E-G}/k^X{}_{E-G}b^X{}_{E-G} + \mathcal{W}_{H-W}/k^X{}_{H-W}b^X{}_{H-W} + \diamondsuit)^* \circ \mathcal{G}$ |
| Force quantum | $O = (\mathcal{W}_{H-S}/k^X{}_{H-S}b^X{}_{H-S} + \mathcal{W}_{S-W}/k^X{}_{S-W}b^X{}_{S-W}$ $+ \mathcal{W}_{E-G}/k^X{}_{E-G}b^X{}_{E-G} + \mathcal{W}_{H-W}/k^X{}_{H-W}b^X{}_{H-W} + \diamondsuit) \circ \mathcal{T}$ |

## 4. Compounding field in trigintaduonion space T-A

It is believed that strong-weak field, hyper-strong field, electromagnetic-gravitational field and hyper-weak field are unified, equal and interconnected. By means of the conception of the space expansion etc., four types of octonionic spaces can be combined into a trigintaduonion space T-A. In trigintaduonion space, some properties of eight sorts of interactions including strong, weak, electromagnetic and gravitational interactions etc. can be described uniformly.

In the trigintaduonion space T-A, there exists one kind of field (trigintaduonion field T-A, for short) which is different to the trigintaduonion field T-X, can be obtained related to the operator ($\mathcal{A}/K + \diamondsuit$). In the trigintaduonion space T-A, the base $\mathcal{E}_{T-A}$ can be written as

$$\mathcal{E}_{T-A} = \mathcal{E}_{T-X} \qquad (34)$$

The displacement $\mathcal{R}_{T-A}$ in trigintaduonion space T-A is

$$\mathcal{R}_{T-A} = \mathcal{R}_{T-X} \qquad (35)$$

The trigintaduonion differential operator $\diamondsuit_{T-A}$ and its conjugate operator are defined as
$$\diamondsuit_{T-A} = \diamondsuit_{T-X} \quad , \quad \diamondsuit^*_{T-A} = \diamondsuit^*_{T-X} \tag{36}$$

In the trigintaduonion field T-A, by analogy with the octonion and sedenion fields, the trigintaduonion differential operator $\diamondsuit$ needs to be generalized to the operator ($\mathcal{A}_{H-S}/k^A_{H-S} + \mathcal{A}_{S-W}/k^A_{S-W} + \mathcal{A}_{E-G}/k^A_{E-G} + \mathcal{A}_{H-W}/k^A_{H-W} + \diamondsuit$). This is because the trigintaduonion field T-A includes hyper-strong, strong-weak, electromagnetic-gravitational and hyper-weak fields.

It can be predicted that the eight sorts of interactions are interconnected each other. The physical features of each subfield in the trigintaduonion field T-A meet the requirements of the equations set in the Table 5.

In the trigintaduonion field T-A, the field potential $\mathcal{A} = (a_0, a_1, a_2, a_3, a_4, a_5, a_6, a_7, a_8, a_9, a_{10}, a_{11}, a_{12}, a_{13}, a_{14}, a_{15}, a_{16}, a_{17}, a_{18}, a_{19}, a_{20}, a_{21}, a_{22}, a_{23}, a_{24}, a_{25}, a_{26}, a_{27}, a_{28}, a_{29}, a_{30}, a_{31})$ is defined as

$$\begin{aligned}\mathcal{A} &= \diamondsuit^* \circ \mathcal{X} \\ &= a_0 + a_1 \vec{e}_1 + a_2 \vec{e}_2 + a_3 \vec{e}_3 + a_4 \vec{e}_4 + a_5 \vec{e}_5 + a_6 \vec{e}_6 \\ &+ a_7 \vec{e}_7 + a_8 \vec{e}_8 + a_9 \vec{e}_9 + a_{10} \vec{e}_{10} + a_{11} \vec{e}_{11} \\ &+ a_{12} \vec{e}_{12} + a_{13} \vec{e}_{13} + a_{14} \vec{e}_{14} + a_{15} \vec{e}_{15} + a_{16} \vec{e}_{16} \\ &+ a_{17} \vec{e}_{17} + a_{18} \vec{e}_{18} + a_{19} \vec{e}_{19} + a_{20} \vec{e}_{20} + a_{21} \vec{e}_{21} \\ &+ a_{22} \vec{e}_{22} + a_{23} \vec{e}_{23} + a_{24} \vec{e}_{24} + a_{25} \vec{e}_{25} + a_{26} \vec{e}_{26} \\ &+ a_{27} \vec{e}_{27} + a_{28} \vec{e}_{28} + a_{29} \vec{e}_{29} + a_{30} \vec{e}_{30} + a_{31} \vec{e}_{31}\end{aligned} \tag{37}$$

where, the mark (*) denotes the trigintaduonion conjugate. $\mathcal{X} = \mathcal{X}_{T-A} = \mathcal{X}_{T-X}$.

The field strength $\mathcal{B}$ of the trigintaduonion field T-A can be defined as
$$\begin{aligned}\mathcal{B} &= (\mathcal{A}/K + \diamondsuit) \circ \mathcal{A} \\ &= (\mathcal{A}_{H-S}/k^A_{H-S} + \mathcal{A}_{S-W}/k^A_{S-W} + \mathcal{A}_{E-G}/k^A_{E-G} + \mathcal{A}_{H-W}/k^A_{H-W} + \diamondsuit) \circ \mathcal{A}\end{aligned} \tag{38}$$

where, $K = K_{T-A}$, $k^A_{H-S}$, $k^A_{S-W}$, $k^A_{E-G}$ and $k^A_{H-W}$ are coefficients in the trigintaduonion space. The field potentials are

$\mathcal{A}_{H-S} = a_0 + a_1 \vec{e}_1 + a_2 \vec{e}_2 + a_3 \vec{e}_3 + a_4 \vec{e}_4 + a_5 \vec{e}_5 + a_6 \vec{e}_6 + a_7 \vec{e}_7$

$\mathcal{A}_{S-W} = a_8 \vec{e}_8 + a_9 \vec{e}_9 + a_{10} \vec{e}_{10} + a_{11} \vec{e}_{11} + a_{12} \vec{e}_{12} + a_{13} \vec{e}_{13} + a_{14} \vec{e}_{14} + a_{15} \vec{e}_{15}$

$\mathcal{A}_{E-G} = a_{16} \vec{e}_{16} + a_{17} \vec{e}_{17} + a_{18} \vec{e}_{18} + a_{19} \vec{e}_{19} + a_{20} \vec{e}_{20} + a_{21} \vec{e}_{21} + a_{22} \vec{e}_{22} + a_{23} \vec{e}_{23}$

$\mathcal{A}_{H-W} = a_{24} \vec{e}_{24} + a_{25} \vec{e}_{25} + a_{26} \vec{e}_{26} + a_{27} \vec{e}_{27} + a_{28} \vec{e}_{28} + a_{29} \vec{e}_{29} + a_{30} \vec{e}_{30} + a_{31} \vec{e}_{31}$

The field source and force of the trigintaduonion field T-A can be defined respectively as
$$\mu \mathcal{S} = (\mathcal{A}_{H-S}/k^A_{H-S} + \mathcal{A}_{S-W}/k^A_{S-W} + \mathcal{A}_{E-G}/k^A_{E-G} + \mathcal{A}_{H-W}/k^A_{H-W} + \diamondsuit)^* \circ \mathcal{B} \tag{39}$$
$$\mathcal{Z} = K (\mathcal{A}_{H-S}/k^A_{H-S} + \mathcal{A}_{S-W}/k^A_{S-W} + \mathcal{A}_{E-G}/k^A_{E-G} + \mathcal{A}_{H-W}/k^A_{H-W} + \diamondsuit) \circ \mathcal{S} \tag{40}$$

where, the coefficient $\mu$ is interaction intensity of the trigintaduonion field T-A.

The angular momentum of trigintaduonion field can be defined as ($k_{rx}$ is the coefficient)
$$\mathcal{M} = \mathcal{S} \circ (\mathcal{r} + k_{rx} \mathcal{X}) \tag{41}$$

and the energy and power in the trigintaduonion field can be defined respectively as
$$\mathcal{W} = K (\mathcal{X}_{H-S}/k^X_{H-S} + \mathcal{X}_{S-W}/k^X_{S-W} + \mathcal{X}_{E-G}/k^X_{E-G} + \mathcal{X}_{H-W}/k^X_{H-W} + \diamondsuit)^* \circ \mathcal{M} \tag{42}$$
$$\mathcal{N} = K (\mathcal{X}_{H-S}/k^X_{H-S} + \mathcal{X}_{S-W}/k^X_{S-W} + \mathcal{X}_{E-G}/k^X_{E-G} + \mathcal{X}_{H-W}/k^X_{H-W} + \diamondsuit) \circ \mathcal{W} \tag{43}$$

In the trigintaduonion space T-A, the wave functions of the quantum mechanics are the trigintaduonion equations set. The Dirac and Klein-Gordon equations of quantum mechanics are actually the wave equations set which are associated with particle's angular momentum.

In the trigintaduonion field T-A, the Dirac equation and the Klein-Gordon equation can be attained respectively from the energy equation (42) and power equation (43) after substituting the operator K ($\mathcal{A}_{H-S} / k^A_{H-S} + \mathcal{A}_{S-W} / k^A_{S-W} + \mathcal{A}_{E-G} / k^A_{E-G} + \mathcal{A}_{H-W} / k^A_{H-W} + \diamond$) for the operator ($\mathcal{W}_{H-S} / k^A_{H-S} b^A_{H-S} + \mathcal{W}_{S-W} / k^A_{S-W} b^A_{S-W} + \mathcal{W}_{E-G} / k^A_{E-G} b^A_{E-G} + \mathcal{W}_{H-W} / k^A_{H-W} b^A_{H-W} + \diamond$). The coefficients $b^A_{H-S}$, $b^A_{S-W}$, $b^A_{E-G}$ and $b^A_{H-W}$ are the Plank-like constant.

Table 5.  Equations set of trigintaduonion field T-A

| Spacetime | trigintaduonion space T-A |
|---|---|
| $\mathcal{X}$ physical quantity | $\mathcal{X} = \mathcal{X}_{T-X}$ |
| Field potential | $\mathcal{A} = \diamond^* \circ \mathcal{X}$ |
| Field strength | $\mathcal{B} = (\mathcal{A}_{H-S} / k^A_{H-S} + \mathcal{A}_{S-W} / k^A_{S-W} + \mathcal{A}_{E-G} / k^A_{E-G} + \mathcal{A}_{H-W} / k^A_{H-W} + \diamond) \circ \mathcal{A}$ |
| Field source | $\mu \mathcal{S} = (\mathcal{A}_{H-S} / k^A_{H-S} + \mathcal{A}_{S-W} / k^A_{S-W} + \mathcal{A}_{E-G} / k^A_{E-G} + \mathcal{A}_{H-W} / k^A_{H-W} + \diamond)^* \circ \mathcal{B}$ |
| Force | $\mathcal{Z} = K (\mathcal{A}_{H-S} / k^A_{H-S} + \mathcal{A}_{S-W} / k^A_{S-W} + \mathcal{A}_{E-G} / k^A_{E-G} + \mathcal{A}_{H-W} / k^A_{H-W} + \diamond) \circ \mathcal{S}$ |
| Angular momentum | $\mathcal{M} = \mathcal{S} \circ (\mathcal{r} + k_{rx} \mathcal{X})$ |
| Energy | $\mathcal{W} = K (\mathcal{A}_{H-S} / k^A_{H-S} + \mathcal{A}_{S-W} / k^A_{S-W} + \mathcal{A}_{E-G} / k^A_{E-G} + \mathcal{A}_{H-W} / k^A_{H-W} + \diamond)^* \circ \mathcal{M}$ |
| Power | $\mathcal{N} = K (\mathcal{A}_{H-S} / k^A_{H-S} + \mathcal{A}_{S-W} / k^A_{S-W} + \mathcal{A}_{E-G} / k^A_{E-G} + \mathcal{A}_{H-W} / k^A_{H-W} + \diamond) \circ \mathcal{W}$ |

The $\mathcal{U}$ equation of the quantum mechanics can be defined as

$$\mathcal{U} = (\mathcal{W}_{H-S} / k^A_{H-S} b^A_{H-S} + \mathcal{W}_{S-W} / k^A_{S-W} b^A_{S-W} + \mathcal{W}_{E-G} / k^A_{E-G} b^A_{E-G} + \mathcal{W}_{H-W} / k^A_{H-W} b^A_{H-W} + \diamond)^* \circ \mathcal{M} \quad (44)$$

The $\mathcal{L}$ equation of the quantum mechanics can be defined as

$$\mathcal{L} = (\mathcal{W}_{H-S} / k^A_{H-S} b^A_{H-S} + \mathcal{W}_{S-W} / k^A_{S-W} b^A_{S-W} + \mathcal{W}_{E-G} / k^A_{E-G} b^A_{E-G} + \mathcal{W}_{H-W} / k^A_{H-W} b^A_{H-W} + \diamond) \circ \mathcal{U} \quad (45)$$

Table 6.  Quantum equations set of trigintaduonion field T-A

| Energy quantum | $\mathcal{U} = (\mathcal{W}_{H-S} / k^A_{H-S} b^A_{H-S} + \mathcal{W}_{S-W} / k^A_{S-W} b^A_{S-W} + \mathcal{W}_{E-G} / k^A_{E-G} b^A_{E-G} + \mathcal{W}_{H-W} / k^A_{H-W} b^A_{H-W} + \diamond)^* \circ \mathcal{M}$ |
|---|---|
| Power quantum | $\mathcal{L} = (\mathcal{W}_{H-S} / k^A_{H-S} b^A_{H-S} + \mathcal{W}_{S-W} / k^A_{S-W} b^A_{S-W} + \mathcal{W}_{E-G} / k^A_{E-G} b^A_{E-G} + \mathcal{W}_{H-W} / k^A_{H-W} b^A_{H-W} + \diamond) \circ \mathcal{U}$ |
| Field strength quantum | $\mathcal{G} = (\mathcal{W}_{H-S} / k^A_{H-S} b^A_{H-S} + \mathcal{W}_{S-W} / k^A_{S-W} b^A_{S-W} + \mathcal{W}_{E-G} / k^A_{E-G} b^A_{E-G} + \mathcal{W}_{H-W} / k^A_{H-W} b^A_{H-W} + \diamond) \circ \mathcal{A}$ |
| Field source quantum | $\mathcal{T} = (\mathcal{W}_{H-S} / k^A_{H-S} b^A_{H-S} + \mathcal{W}_{S-W} / k^A_{S-W} b^A_{S-W} + \mathcal{W}_{E-G} / k^A_{E-G} b^A_{E-G} + \mathcal{W}_{H-W} / k^A_{H-W} b^A_{H-W} + \diamond)^* \circ \mathcal{G}$ |
| Force quantum | $\mathcal{O} = (\mathcal{W}_{H-S} / k^A_{H-S} b^A_{H-S} + \mathcal{W}_{S-W} / k^A_{S-W} b^A_{S-W} + \mathcal{W}_{E-G} / k^A_{E-G} b^A_{E-G} + \mathcal{W}_{H-W} / k^A_{H-W} b^A_{H-W} + \diamond) \circ \mathcal{T}$ |

The three sorts of Dirac-like equations can be obtained from Eqs.(38), (39) and (40) respectively.

The $\mathcal{G}$ equation of the quantum mechanics can be defined as

$$\mathcal{G} = (\mathcal{W}_{H-S} / k^A_{H-S} b^A_{H-S} + \mathcal{W}_{S-W} / k^A_{S-W} b^A_{S-W} + \mathcal{W}_{E-G} / k^A_{E-G} b^A_{E-G} + \mathcal{W}_{H-W} / k^A_{H-W} b^A_{H-W} + \diamond) \circ \mathcal{A} \quad (46)$$

The $\mathcal{T}$ equation of the quantum mechanics can be defined as

$$\mathcal{T} = (\mathcal{W}_{H-S} / k^A_{H-S} b^A_{H-S} + \mathcal{W}_{S-W} / k^A_{S-W} b^A_{S-W}$$

$$+ \mathcal{W}_{E-G} / k^A_{E-G} b^A_{E-G} + \mathcal{W}_{H-W} / k^A_{H-W} b^A_{H-W} + \diamondsuit)^* \circ \mathcal{G} \tag{47}$$

The $O$ equation of the quantum mechanics can be defined as

$$O = (\mathcal{W}_{H-S} / k^A_{H-S} b^A_{H-S} + \mathcal{W}_{S-W} / k^A_{S-W} b^A_{S-W}$$
$$+ \mathcal{W}_{E-G} / k^A_{E-G} b^A_{E-G} + \mathcal{W}_{H-W} / k^A_{H-W} b^A_{H-W} + \diamondsuit) \circ \mathcal{T} \tag{48}$$

In the trigintaduonion field T-A, the intermediate and field source particles can be obtained. We can find that the intermediate particles and other kinds of new and unknown particles may be existed in the nature.

## 5. Compounding field in trigintaduonion space T-B

It is believed that electromagnetic-gravitational field, strong-weak field, hyper-strong field and hyper-weak field are unified, equal and interconnected. By means of the conception of the space expansion etc., four types of octonionic spaces can be combined into a trigintaduonion space T-B. In trigintaduonion space, some properties of eight sorts of interactions including strong, weak, electromagnetic and gravitational interactions etc. can be described uniformly.

In the trigintaduonion space T-B, there exists one kind of field (trigintaduonion field T-B, for short) which is different to the trigintaduonion field T-X or T-A, can be obtained related to the operator ($\mathcal{B}/K + \diamondsuit$). In the trigintaduonion space T-B, the base $\mathcal{E}_{T-B}$ can be written as

$$\mathcal{E}_{T-B} = \mathcal{E}_{T-X} \tag{49}$$

The displacement $\mathcal{R}_{T-B}$ in trigintaduonion space T-B is

$$\mathcal{R}_{T-B} = \mathcal{R}_{T-X} \tag{50}$$

The trigintaduonion differential operator $\diamondsuit_{T-B}$ and its conjugate operator are defined as

$$\diamondsuit_{T-B} = \diamondsuit_{T-X} \quad , \quad \diamondsuit^*_{T-B} = \diamondsuit^*_{T-X} \tag{51}$$

In the trigintaduonion field T-B, by analogy with the octonion and sedenion fields, the trigintaduonion differential operator $\diamondsuit$ needs to be generalized to the operator ($\mathcal{B}_{H-S} / k^B_{H-S} + \mathcal{B}_{S-W} / k^B_{S-W} + \mathcal{B}_{E-G} / k^B_{E-G} + \mathcal{B}_{H-W} / k^B_{H-W} + \diamondsuit$). This is because the trigintaduonion field T-B includes hyper-strong, strong-weak, electromagnetic-gravitational and hyper-weak fields.

It can be predicted that the eight sorts of interactions are interconnected each other. The physical features of each subfield in the trigintaduonion field T-B meet the requirements of the equations set in the Table 7.

In the trigintaduonion field T-B, the field potential $\mathcal{A} = (a_0, a_1, a_2, a_3, a_4, a_5, a_6, a_7, a_8, a_9, a_{10}, a_{11}, a_{12}, a_{13}, a_{14}, a_{15}, a_{16}, a_{17}, a_{18}, a_{19}, a_{20}, a_{21}, a_{22}, a_{23}, a_{24}, a_{25}, a_{26}, a_{27}, a_{28}, a_{29}, a_{30}, a_{31})$ is defined as

$$\mathcal{A} = \diamondsuit^* \circ \mathcal{X}$$
$$= a_0 + a_1 \vec{e}_1 + a_2 \vec{e}_2 + a_3 \vec{e}_3 + a_4 \vec{e}_4 + a_5 \vec{e}_5 + a_6 \vec{e}_6$$
$$+ a_7 \vec{e}_7 + a_8 \vec{e}_8 + a_9 \vec{e}_9 + a_{10} \vec{e}_{10} + a_{11} \vec{e}_{11}$$
$$+ a_{12} \vec{e}_{12} + a_{13} \vec{e}_{13} + a_{14} \vec{e}_{14} + a_{15} \vec{e}_{15} + a_{16} \vec{e}_{16}$$
$$+ a_{17} \vec{e}_{17} + a_{18} \vec{e}_{18} + a_{19} \vec{e}_{19} + a_{20} \vec{e}_{20} + a_{21} \vec{e}_{21}$$
$$+ a_{22} \vec{e}_{22} + a_{23} \vec{e}_{23} + a_{24} \vec{e}_{24} + a_{25} \vec{e}_{25} + a_{26} \vec{e}_{26}$$
$$+ a_{27} \vec{e}_{27} + a_{28} \vec{e}_{28} + a_{29} \vec{e}_{29} + a_{30} \vec{e}_{30} + a_{31} \vec{e}_{31} \tag{52}$$

where, the mark (*) denotes the trigintaduonion conjugate. $\mathcal{X} = \mathcal{X}_{T-B} = \mathcal{X}_{T-X}$.

The field strength $\mathcal{B}$ of the trigintaduonion field T-B can be defined as

$$\mathcal{B} = \diamondsuit \circ \mathcal{A} \tag{53}$$

The field source of the trigintaduonion field T-B can be defined as

$$\mu S = (\mathcal{B}/K + \diamondsuit)^* \circ \mathcal{B}$$
$$= (\mathcal{B}_{H-S} / k^B_{H-S} + \mathcal{B}_{S-W} / k^B_{S-W} + \mathcal{B}_{E-G} / k^B_{E-G} + \mathcal{B}_{H-W} / k^B_{H-W} + \diamondsuit)^* \circ \mathcal{B} \quad (54)$$

where, $K = K_{T-B}$, $k^B_{H-S}$, $k^B_{S-W}$, $k^B_{E-G}$ and $k^B_{H-W}$ are coefficients in the trigintaduonion space. The coefficient $\mu$ is interaction intensity of trigintaduonion field T-B. The field strengths are

$\mathcal{B}_{H-S} = b_0 + b_1 \vec{e}_1 + b_2 \vec{e}_2 + b_3 \vec{e}_3 + b_4 \vec{e}_4 + b_5 \vec{e}_5 + b_6 \vec{e}_6 + b_7 \vec{e}_7$

$\mathcal{B}_{S-W} = b_8 \vec{e}_8 + b_9 \vec{e}_9 + b_{10} \vec{e}_{10} + b_{11} \vec{e}_{11} + b_{12} \vec{e}_{12} + b_{13} \vec{e}_{13} + b_{14} \vec{e}_{14} + b_{15} \vec{e}_{15}$

$\mathcal{B}_{E-G} = b_{16} \vec{e}_{16} + b_{17} \vec{e}_{17} + b_{18} \vec{e}_{18} + b_{19} \vec{e}_{19} + b_{20} \vec{e}_{20} + b_{21} \vec{e}_{21} + b_{22} \vec{e}_{22} + b_{23} \vec{e}_{23}$

$\mathcal{B}_{H-W} = b_{24} \vec{e}_{24} + b_{25} \vec{e}_{25} + b_{26} \vec{e}_{26} + b_{27} \vec{e}_{27} + b_{28} \vec{e}_{28} + b_{29} \vec{e}_{29} + b_{30} \vec{e}_{30} + b_{31} \vec{e}_{31}$

The force of the trigintaduonion field T-B can be defined as

$$Z = K (\mathcal{B}_{H-S} / k^B_{H-S} + \mathcal{B}_{S-W} / k^B_{S-W} + \mathcal{B}_{E-G} / k^B_{E-G} + \mathcal{B}_{H-W} / k^B_{H-W} + \diamondsuit) \circ S \quad (55)$$

The angular momentum of trigintaduonion field can be defined as ($k_{rx}$ is the coefficient)

$$\mathcal{M} = S \circ (r + k_{rx} X) \quad (56)$$

and the energy and power in the trigintaduonion field can be defined respectively as

$$\mathcal{W} = K (\mathcal{B}_{H-S} / k^B_{H-S} + \mathcal{B}_{S-W} / k^B_{S-W} + \mathcal{B}_{E-G} / k^B_{E-G} + \mathcal{B}_{H-W} / k^B_{H-W} + \diamondsuit)^* \circ \mathcal{M} \quad (57)$$

$$\mathcal{N} = K (\mathcal{B}_{H-S} / k^B_{H-S} + \mathcal{B}_{S-W} / k^B_{S-W} + \mathcal{B}_{E-G} / k^B_{E-G} + \mathcal{B}_{H-W} / k^B_{H-W} + \diamondsuit) \circ \mathcal{W} \quad (58)$$

In the trigintaduonion space T-B, the wave functions of the quantum mechanics are the trigintaduonion equations set. The Dirac and Klein-Gordon equations of quantum mechanics are actually the wave equations set which are associated with particle's angular momentum.

In the trigintaduonion field T-B, the Dirac equation and the Klein-Gordon equation can be attained respectively from the energy equation (57) and power equation (58) after substituting the operator $K (\mathcal{B}_{H-S} / k^B_{H-S} + \mathcal{B}_{S-W} / k^B_{S-W} + \mathcal{B}_{E-G} / k^B_{E-G} + \mathcal{B}_{H-W} / k^B_{H-W} + \diamondsuit)$ for the operator $(\mathcal{W}_{H-S} / k^B_{H-S} b^B_{H-S} + \mathcal{W}_{S-W} / k^B_{S-W} b^B_{S-W} + \mathcal{W}_{E-G} / k^B_{E-G} b^B_{E-G} + \mathcal{W}_{H-W} / k^B_{H-W} b^B_{H-W} + \diamondsuit)$. The coefficients $b^B_{H-S}$, $b^B_{S-W}$, $b^B_{E-G}$ and $b^B_{H-W}$ are the Plank-like constant.

Table 7.   Equations set of trigintaduonion field T-B

| Spacetime | trigintaduonion space T-B |
|---|---|
| $X$ physical quantity | $X = X_{T-X}$ |
| Field potential | $\mathcal{A} = \diamondsuit^* \circ X$ |
| Field strength | $\mathcal{B} = \diamondsuit \circ \mathcal{A}$ |
| Field source | $\mu S = (\mathcal{B}_{H-S} / k^B_{H-S} + \mathcal{B}_{S-W} / k^B_{S-W} + \mathcal{B}_{E-G} / k^B_{E-G} + \mathcal{B}_{H-W} / k^B_{H-W} + \diamondsuit)^* \circ \mathcal{B}$ |
| Force | $Z = K (\mathcal{B}_{H-S} / k^B_{H-S} + \mathcal{B}_{S-W} / k^B_{S-W} + \mathcal{B}_{E-G} / k^B_{E-G} + \mathcal{B}_{H-W} / k^B_{H-W} + \diamondsuit) \circ S$ |
| Angular momentum | $\mathcal{M} = S \circ (r + k_{rx} X)$ |
| Energy | $\mathcal{W} = K (\mathcal{B}_{H-S} / k^B_{H-S} + \mathcal{B}_{S-W} / k^B_{S-W} + \mathcal{B}_{E-G} / k^B_{E-G} + \mathcal{B}_{H-W} / k^B_{H-W} + \diamondsuit)^* \circ \mathcal{M}$ |
| Power | $\mathcal{N} = K (\mathcal{B}_{H-S} / k^B_{H-S} + \mathcal{B}_{S-W} / k^B_{S-W} + \mathcal{B}_{E-G} / k^B_{E-G} + \mathcal{B}_{H-W} / k^B_{H-W} + \diamondsuit) \circ \mathcal{W}$ |

The $U$ equation of the quantum mechanics can be defined as

$$U = (\mathcal{W}_{H-S} / k^B_{H-S} b^B_{H-S} + \mathcal{W}_{S-W} / k^B_{S-W} b^B_{S-W} + \mathcal{W}_{E-G} / k^B_{E-G} b^B_{E-G} + \mathcal{W}_{H-W} / k^B_{H-W} b^B_{H-W} + \diamondsuit)^* \circ \mathcal{M} \quad (59)$$

The $\mathcal{L}$ equation of the quantum mechanics can be defined as

$$\mathcal{L} = (\mathcal{W}_{H-S} / k^B_{H-S} b^B_{H-S} + \mathcal{W}_{S-W} / k^B_{S-W} b^B_{S-W} + \mathcal{W}_{E-G} / k^B_{E-G} b^B_{E-G} + \mathcal{W}_{H-W} / k^B_{H-W} b^B_{H-W} + \diamondsuit) \circ U \quad (60)$$

The two sorts of Dirac-like equations can be obtained from the field source equation (54) and force equation (55) respectively.

The $\mathcal{T}$ equation of the quantum mechanics can be defined as

$$\mathcal{T} = (\mathcal{W}_{H-S} / k^B_{H-S} b^B_{H-S} + \mathcal{W}_{S-W} / k^B_{S-W} b^B_{S-W}$$
$$+ \mathcal{W}_{E-G} / k^B_{E-G} b^B_{E-G} + \mathcal{W}_{H-W} / k^B_{H-W} b^B_{H-W} + \diamondsuit)^* \circ \mathcal{B} \quad (61)$$

The $O$ equation of the quantum mechanics can be defined as

$$O = (\mathcal{W}_{H-S} / k^B_{H-S} b^B_{H-S} + \mathcal{W}_{S-W} / k^B_{S-W} b^B_{S-W}$$
$$+ \mathcal{W}_{E-G} / k^B_{E-G} b^B_{E-G} + \mathcal{W}_{H-W} / k^B_{H-W} b^B_{H-W} + \diamondsuit) \circ \mathcal{T} \quad (62)$$

Table 8. Quantum equations set of trigintaduonion field T-B

| Energy quantum | $\mathcal{U} = (\mathcal{W}_{H-S} / k^B_{H-S} b^B_{H-S} + \mathcal{W}_{S-W} / k^B_{S-W} b^B_{S-W} + \mathcal{W}_{E-G} / k^B_{E-G} b^B_{E-G} + \mathcal{W}_{H-W} / k^B_{H-W} b^B_{H-W} + \diamondsuit)^* \circ \mathcal{M}$ |
|---|---|
| Power quantum | $\mathcal{L} = (\mathcal{W}_{H-S} / k^B_{H-S} b^B_{H-S} + \mathcal{W}_{S-W} / k^B_{S-W} b^B_{S-W} + \mathcal{W}_{E-G} / k^B_{E-G} b^B_{E-G} + \mathcal{W}_{H-W} / k^B_{H-W} b^B_{H-W} + \diamondsuit) \circ \mathcal{U}$ |
| Field source quantum | $\mathcal{T} = (\mathcal{W}_{H-S} / k^B_{H-S} b^B_{H-S} + \mathcal{W}_{S-W} / k^B_{S-W} b^B_{S-W} + \mathcal{W}_{E-G} / k^B_{E-G} b^B_{E-G} + \mathcal{W}_{H-W} / k^B_{H-W} b^B_{H-W} + \diamondsuit)^* \circ \mathcal{B}$ |
| Force quantum | $O = (\mathcal{W}_{H-S} / k^B_{H-S} b^B_{H-S} + \mathcal{W}_{S-W} / k^B_{S-W} b^B_{S-W} + \mathcal{W}_{E-G} / k^B_{E-G} b^B_{E-G} + \mathcal{W}_{H-W} / k^B_{H-W} b^B_{H-W} + \diamondsuit) \circ \mathcal{T}$ |

In the trigintaduonion field T-B, the intermediate and field source particles can be obtained. We can find that the intermediate particles and other kinds of new and unknown particles may be existed in the nature.

## 6. Compounding field in trigintaduonion space T-S

It is believed that hyper-weak field, electromagnetic-gravitational field, strong-weak field and hyper-strong field are unified, equal and interconnected. By means of the conception of the space expansion etc., four types of octonionic spaces can be combined into a trigintaduonion space T-S. In trigintaduonion space, some properties of eight sorts of interactions including strong, weak, electromagnetic and gravitational interactions etc. can be described uniformly.

In the trigintaduonion space T-S, there exists one kind of field (trigintaduonion field T-S, for short) which is different to trigintaduonion field T-X, T-A or T-B, can be obtained related to operator ($S/K + \diamondsuit$). In the trigintaduonion space T-S, the base $\mathcal{E}_{T-S}$ can be written as

$$\mathcal{E}_{T-S} = \mathcal{E}_{T-X} \quad (63)$$

The displacement $\mathcal{R}_{T-S}$ in trigintaduonion space T-S is

$$\mathcal{R}_{T-S} = \mathcal{R}_{T-X} \quad (64)$$

The trigintaduonion differential operator $\diamondsuit_{T-S}$ and its conjugate operator are defined as

$$\diamondsuit_{T-S} = \diamondsuit_{T-X}, \quad \diamondsuit^*_{T-S} = \diamondsuit^*_{T-X} \quad (65)$$

In the trigintaduonion field T-S, by analogy with the octonion and sedenion fields, the trigintaduonion differential operator $\diamondsuit$ needs to be generalized to a new operator ($S_{H-S} / k^S_{H-S} + S_{S-W} / k^S_{S-W} + S_{E-G} / k^S_{E-G} + S_{H-W} / k^S_{H-W} + \diamondsuit$). This is because the trigintaduonion field T-S includes the hyper-strong, strong-weak, electromagnetic-gravitational and hyper-weak fields.

It can be predicted that the eight sorts of interactions are interconnected each other. The physical features of each subfield in the trigintaduonion field T-S meet the requirements of the equations set in the Table 9.

In the trigintaduonion field T-S, the field potential $\mathcal{A} = (a_0, a_1, a_2, a_3, a_4, a_5, a_6, a_7, a_8, a_9, a_{10}, a_{11}, a_{12}, a_{13}, a_{14}, a_{15}, a_{16}, a_{17}, a_{18}, a_{19}, a_{20}, a_{21}, a_{22}, a_{23}, a_{24}, a_{25}, a_{26}, a_{27}, a_{28}, a_{29}, a_{30}, a_{31})$ is defined as

$$\mathcal{A} = \Diamond^* \circ \mathcal{X}$$
$$= a_0 + a_1 \vec{e}_1 + a_2 \vec{e}_2 + a_3 \vec{e}_3 + a_4 \vec{e}_4 + a_5 \vec{e}_5 + a_6 \vec{e}_6$$
$$+ a_7 \vec{e}_7 + a_8 \vec{e}_8 + a_9 \vec{e}_9 + a_{10} \vec{e}_{10} + a_{11} \vec{e}_{11}$$
$$+ a_{12} \vec{e}_{12} + a_{13} \vec{e}_{13} + a_{14} \vec{e}_{14} + a_{15} \vec{e}_{15} + a_{16} \vec{e}_{16}$$
$$+ a_{17} \vec{e}_{17} + a_{18} \vec{e}_{18} + a_{19} \vec{e}_{19} + a_{20} \vec{e}_{20} + a_{21} \vec{e}_{21}$$
$$+ a_{22} \vec{e}_{22} + a_{23} \vec{e}_{23} + a_{24} \vec{e}_{24} + a_{25} \vec{e}_{25} + a_{26} \vec{e}_{26}$$
$$+ a_{27} \vec{e}_{27} + a_{28} \vec{e}_{28} + a_{29} \vec{e}_{29} + a_{30} \vec{e}_{30} + a_{31} \vec{e}_{31} \tag{66}$$

where, the mark (*) denotes the trigintaduonion conjugate. $\mathcal{X} = \mathcal{X}_{T-S} = \mathcal{X}_{T-X}$.

The field strength $\mathcal{B}$ of the trigintaduonion field T-S can be defined as

$$\mathcal{B} = \Diamond \circ \mathcal{A} \tag{67}$$

The field source of the trigintaduonion field T-S can be defined as

$$\mu \mathcal{S} = \Diamond^* \circ \mathcal{B} \tag{68}$$

where, the coefficient $\mu$ is interaction intensity of the trigintaduonion field T-S.

The force of the trigintaduonion field T-S can be defined as

$$\mathcal{Z} = \mathbb{K}\,(\mathcal{S}_{H-S}/k^S_{H-S} + \mathcal{S}_{S-W}/k^S_{S-W} + \mathcal{S}_{E-G}/k^S_{E-G} + \mathcal{S}_{H-W}/k^S_{H-W} + \Diamond) \circ \mathcal{S} \tag{69}$$

where, $\mathbb{K} = \mathbb{K}_{T-S}$, $k^S_{H-S}$, $k^S_{S-W}$, $k^S_{E-G}$ and $k^S_{H-W}$ are coefficients in the trigintaduonion space. And the field sources are

$\mathcal{S}_{H-S} = s_0 + s_1 \vec{e}_1 + s_2 \vec{e}_2 + s_3 \vec{e}_3 + s_4 \vec{e}_4 + s_5 \vec{e}_5 + s_6 \vec{e}_6 + s_7 \vec{e}_7$

$\mathcal{S}_{S-W} = s_8 \vec{e}_8 + s_9 \vec{e}_9 + s_{10} \vec{e}_{10} + s_{11} \vec{e}_{11} + s_{12} \vec{e}_{12} + s_{13} \vec{e}_{13} + s_{14} \vec{e}_{14} + s_{15} \vec{e}_{15}$

$\mathcal{S}_{E-G} = s_{16} \vec{e}_{16} + s_{17} \vec{e}_{17} + s_{18} \vec{e}_{18} + s_{19} \vec{e}_{19} + s_{20} \vec{e}_{20} + s_{21} \vec{e}_{21} + s_{22} \vec{e}_{22} + s_{23} \vec{e}_{23}$

$\mathcal{S}_{H-W} = s_{24} \vec{e}_{24} + s_{25} \vec{e}_{25} + s_{26} \vec{e}_{26} + s_{27} \vec{e}_{27} + s_{28} \vec{e}_{28} + s_{29} \vec{e}_{29} + s_{30} \vec{e}_{30} + s_{31} \vec{e}_{31}$

Table 9.   Equations set of trigintaduonion field T-S

| Spacetime | trigintaduonion space T-S |
|---|---|
| $\mathcal{X}$ physical quantity | $\mathcal{X} = \mathcal{X}_{T-X}$ |
| Field potential | $\mathcal{A} = \Diamond^* \circ \mathcal{X}$ |
| Field strength | $\mathcal{B} = \Diamond \circ \mathcal{A}$ |
| Field source | $\mu \mathcal{S} = \Diamond^* \circ \mathcal{B}$ |
| Force | $\mathcal{Z} = \mathbb{K}\,(\mathcal{S}_{H-S}/k^S_{H-S} + \mathcal{S}_{S-W}/k^S_{S-W} + \mathcal{S}_{E-G}/k^S_{E-G} + \mathcal{S}_{H-W}/k^S_{H-W} + \Diamond) \circ \mathcal{S}$ |
| Angular momentum | $\mathcal{M} = \mathcal{S} \circ (\mathcal{r} + k_{rx} \mathcal{X})$ |
| Energy | $\mathcal{W} = \mathbb{K}\,(\mathcal{S}_{H-S}/k^S_{H-S} + \mathcal{S}_{S-W}/k^S_{S-W} + \mathcal{S}_{E-G}/k^S_{E-G} + \mathcal{S}_{H-W}/k^S_{H-W} + \Diamond)^* \circ \mathcal{M}$ |
| Power | $\mathcal{N} = \mathbb{K}\,(\mathcal{S}_{H-S}/k^S_{H-S} + \mathcal{S}_{S-W}/k^S_{S-W} + \mathcal{S}_{E-G}/k^S_{E-G} + \mathcal{S}_{H-W}/k^S_{H-W} + \Diamond) \circ \mathcal{W}$ |

The angular momentum of trigintaduonion field can be defined as ($k_{rx}$ is the coefficient)

$$\mathcal{M} = \mathcal{S} \circ (\mathcal{r} + k_{rx} \mathcal{X}) \tag{70}$$

and the energy and power in the trigintaduonion field can be defined respectively as

$$\mathcal{W} = \mathrm{K} \, (S_{\text{H-S}} / \mathrm{k}^S_{\text{H-S}} + S_{\text{S-W}} / \mathrm{k}^S_{\text{S-W}} + S_{\text{E-G}} / \mathrm{k}^S_{\text{E-G}} + S_{\text{H-W}} / \mathrm{k}^S_{\text{H-W}} + \diamondsuit)^* \circ \mathcal{M} \quad (71)$$

$$\mathcal{N} = \mathrm{K} \, (S_{\text{H-S}} / \mathrm{k}^S_{\text{H-S}} + S_{\text{S-W}} / \mathrm{k}^S_{\text{S-W}} + S_{\text{E-G}} / \mathrm{k}^S_{\text{E-G}} + S_{\text{H-W}} / \mathrm{k}^S_{\text{H-W}} + \diamondsuit) \circ \mathcal{W} \quad (72)$$

In the trigintaduonion space T-S, the wave functions of the quantum mechanics are the trigintaduonion equations set. The Dirac and Klein-Gordon equations of quantum mechanics are actually the wave equations set which are associated with particle's angular momentum.

In the trigintaduonion field T-S, the Dirac equation and the Klein-Gordon equation can be attained respectively from the energy equation (71) and power equation (72) after substituting the operator $\mathrm{K} \, (S_{\text{H-S}} / \mathrm{k}^S_{\text{H-S}} + S_{\text{S-W}} / \mathrm{k}^S_{\text{S-W}} + S_{\text{E-G}} / \mathrm{k}^S_{\text{E-G}} + S_{\text{H-W}} / \mathrm{k}^S_{\text{H-W}} + \diamondsuit)$ for the new operator $(\mathcal{W}_{\text{H-S}} / \mathrm{k}^S_{\text{H-S}} \mathrm{b}^S_{\text{H-S}} + \mathcal{W}_{\text{S-W}} / \mathrm{k}^S_{\text{S-W}} \mathrm{b}^S_{\text{S-W}} + \mathcal{W}_{\text{E-G}} / \mathrm{k}^S_{\text{E-G}} \mathrm{b}^S_{\text{E-G}} + \mathcal{W}_{\text{H-W}} / \mathrm{k}^S_{\text{H-W}} \mathrm{b}^S_{\text{H-W}} + \diamondsuit)$. The coefficients $\mathrm{b}^S_{\text{H-S}}$, $\mathrm{b}^S_{\text{S-W}}$, $\mathrm{b}^S_{\text{E-G}}$ and $\mathrm{b}^S_{\text{H-W}}$ are the Plank-like constant.

The $\mathcal{U}$ equation of the quantum mechanics can be defined as

$$\mathcal{U} = (\mathcal{W}_{\text{H-S}} / \mathrm{k}^S_{\text{H-S}} \mathrm{b}^S_{\text{H-S}} + \mathcal{W}_{\text{S-W}} / \mathrm{k}^S_{\text{S-W}} \mathrm{b}^S_{\text{S-W}} + \mathcal{W}_{\text{E-G}} / \mathrm{k}^S_{\text{E-G}} \mathrm{b}^S_{\text{E-G}} + \mathcal{W}_{\text{H-W}} / \mathrm{k}^S_{\text{H-W}} \mathrm{b}^S_{\text{H-W}} + \diamondsuit)^* \circ \mathcal{M} \quad (73)$$

The $\mathcal{L}$ equation of the quantum mechanics can be defined as

$$\mathcal{L} = (\mathcal{W}_{\text{H-S}} / \mathrm{k}^S_{\text{H-S}} \mathrm{b}^S_{\text{H-S}} + \mathcal{W}_{\text{S-W}} / \mathrm{k}^S_{\text{S-W}} \mathrm{b}^S_{\text{S-W}} + \mathcal{W}_{\text{E-G}} / \mathrm{k}^S_{\text{E-G}} \mathrm{b}^S_{\text{E-G}} + \mathcal{W}_{\text{H-W}} / \mathrm{k}^S_{\text{H-W}} \mathrm{b}^S_{\text{H-W}} + \diamondsuit) \circ \mathcal{U} \quad (74)$$

The Dirac-like equation can be obtained from the force equation (69). The $O$ equation of the quantum mechanics can be defined as

$$O = (\mathcal{W}_{\text{H-S}} / \mathrm{k}^S_{\text{H-S}} \mathrm{b}^S_{\text{H-S}} + \mathcal{W}_{\text{S-W}} / \mathrm{k}^S_{\text{S-W}} \mathrm{b}^S_{\text{S-W}} + \mathcal{W}_{\text{E-G}} / \mathrm{k}^S_{\text{E-G}} \mathrm{b}^S_{\text{E-G}} + \mathcal{W}_{\text{H-W}} / \mathrm{k}^S_{\text{H-W}} \mathrm{b}^S_{\text{H-W}} + \diamondsuit) \circ S \quad (75)$$

Table 10. Quantum equations set of trigintaduonion field T-S

| | |
|---|---|
| Energy quantum | $\mathcal{U} = (\mathcal{W}_{\text{H-S}} / \mathrm{k}^S_{\text{H-S}} \mathrm{b}^S_{\text{H-S}} + \mathcal{W}_{\text{S-W}} / \mathrm{k}^S_{\text{S-W}} \mathrm{b}^S_{\text{S-W}} + \mathcal{W}_{\text{E-G}} / \mathrm{k}^S_{\text{E-G}} \mathrm{b}^S_{\text{E-G}} + \mathcal{W}_{\text{H-W}} / \mathrm{k}^S_{\text{H-W}} \mathrm{b}^S_{\text{H-W}} + \diamondsuit)^* \circ \mathcal{M}$ |
| Power quantum | $\mathcal{L} = (\mathcal{W}_{\text{H-S}} / \mathrm{k}^S_{\text{H-S}} \mathrm{b}^S_{\text{H-S}} + \mathcal{W}_{\text{S-W}} / \mathrm{k}^S_{\text{S-W}} \mathrm{b}^S_{\text{S-W}} + \mathcal{W}_{\text{E-G}} / \mathrm{k}^S_{\text{E-G}} \mathrm{b}^S_{\text{E-G}} + \mathcal{W}_{\text{H-W}} / \mathrm{k}^S_{\text{H-W}} \mathrm{b}^S_{\text{H-W}} + \diamondsuit) \circ \mathcal{U}$ |
| Force quantum | $O = (\mathcal{W}_{\text{H-S}} / \mathrm{k}^S_{\text{H-S}} \mathrm{b}^S_{\text{H-S}} + \mathcal{W}_{\text{S-W}} / \mathrm{k}^S_{\text{S-W}} \mathrm{b}^S_{\text{S-W}} + \mathcal{W}_{\text{E-G}} / \mathrm{k}^S_{\text{E-G}} \mathrm{b}^S_{\text{E-G}} + \mathcal{W}_{\text{H-W}} / \mathrm{k}^S_{\text{H-W}} \mathrm{b}^S_{\text{H-W}} + \diamondsuit) \circ S$ |

In the trigintaduonion field T-S, the intermediate and field source particles can be obtained. We can find that the intermediate particles and other kinds of new and unknown particles may be existed in the nature.

## 7. Special case of compounding field in trigintaduonion space

It is believed that different sorts of interactions are all unified, equal and interconnected. By means of the conception of the space expansion etc., four types of the octonionic spaces can be combined into a trigintaduonion space T-C. In the trigintaduonion space, some properties of eight sorts of interactions including the strong, weak, electromagnetic and gravitational interactions etc. can be described uniformly.

In the trigintaduonion space T-C, there exists one kind of field (trigintaduonion field T-C, for short) which is the special case of the trigintaduonion fields T-X, T-A, T-B or T-S, can be obtained related to the operator $\diamondsuit$.

In the trigintaduonion space T-C, the base $\mathcal{E}_{T-C}$ can be written as
$$\mathcal{E}_{T-C} = \mathcal{E}_{T-X} \tag{76}$$

The displacement $\mathcal{R}_{T-C}$ in trigintaduonion space T-C is
$$\mathcal{R}_{T-C} = \mathcal{R}_{T-X} \tag{77}$$

The trigintaduonion differential operator $\diamondsuit_{T-C}$ and its conjugate operator are defined as
$$\diamondsuit_{T-C} = \diamondsuit_{T-X} \ , \quad \diamondsuit^*_{T-C} = \diamondsuit^*_{T-X} \tag{78}$$

It can be predicted that the eight sorts of interactions are interconnected each other. The physical features of each subfield in the trigintaduonion field T-C meet the requirements of the equations set in the Table 11.

In the trigintaduonion field T-C, the field potential $\mathcal{A} = (a_0, a_1, a_2, a_3, a_4, a_5, a_6, a_7, a_8, a_9, a_{10}, a_{11}, a_{12}, a_{13}, a_{14}, a_{15}, a_{16}, a_{17}, a_{18}, a_{19}, a_{20}, a_{21}, a_{22}, a_{23}, a_{24}, a_{25}, a_{26}, a_{27}, a_{28}, a_{29}, a_{30}, a_{31})$ is defined as
$$\mathcal{A} = \diamondsuit^* \circ \mathcal{X} \tag{79}$$

where, the mark (*) denotes the trigintaduonion conjugate. $\mathcal{X} = \mathcal{X}_{T-C} = \mathcal{X}_{T-X}$.

The field strength $\mathcal{B}$ of the trigintaduonion field T-C can be defined as
$$\mathcal{B} = \diamondsuit \circ \mathcal{A} \tag{80}$$

The field source of the trigintaduonion field T-C can be defined as
$$\mu \mathcal{S} = \diamondsuit^* \circ \mathcal{B} \tag{81}$$

where, the coefficient $\mu$ is interaction intensity of the trigintaduonion field T-C.

The force of the trigintaduonion field T-C can be defined as
$$\mathcal{Z} = K \diamondsuit \circ \mathcal{S} \tag{82}$$

where, $K = K_{T-C}$ is the coefficient in the trigintaduonion space.

The angular momentum of trigintaduonion field can be defined as ($k_{rx}$ is the coefficient)
$$\mathcal{M} = \mathcal{S} \circ (\mathcal{R} + k_{rx} \mathcal{X}) \tag{83}$$

and the energy and power in the trigintaduonion field can be defined respectively as
$$\mathcal{W} = K \diamondsuit^* \circ \mathcal{M} \tag{84}$$
$$\mathcal{N} = K \diamondsuit \circ \mathcal{W} \tag{85}$$

Table 11.　Equations set of trigintaduonion field T-C

| Spacetime | trigintaduonion space T-C |
| --- | --- |
| $\mathcal{X}$ physical quantity | $\mathcal{X} = \mathcal{X}_{T-X}$ |
| Field potential | $\mathcal{A} = \diamondsuit^* \circ \mathcal{X}$ |
| Field strength | $\mathcal{B} = \diamondsuit \circ \mathcal{A}$ |
| Field source | $\mu \mathcal{S} = \diamondsuit^* \circ \mathcal{B}$ |
| Force | $\mathcal{Z} = K \diamondsuit \circ \mathcal{S}$ |
| Angular momentum | $\mathcal{M} = \mathcal{S} \circ (\mathcal{R} + k_{rx} \mathcal{X})$ |
| Energy | $\mathcal{W} = K \diamondsuit^* \circ \mathcal{M}$ |
| Power | $\mathcal{N} = K \diamondsuit \circ \mathcal{W}$ |

In the trigintaduonion space T-C, the wave functions of the quantum mechanics are the trigintaduonion equations set. The Dirac and Klein-Gordon equations of quantum mechanics are actually the wave equations set which are associated with particle's angular momentum $\mathcal{M} = b\Psi$. The coefficient b is the Plank-like constant.

In the trigintaduonion field T-C, the Dirac equation and the Klein-Gordon equation can be attained respectively from the energy equation (84) and the power equation (85).

The $\mathcal{U}$ equation of the quantum mechanics can be defined as
$$\mathcal{U} = (b\diamond)^* \circ (\mathcal{M} / b) \tag{86}$$

The $\mathcal{L}$ equation of the quantum mechanics can be defined as
$$\mathcal{L} = (b\diamond) \circ (\mathcal{U} / b) \tag{87}$$

The four sorts of Dirac-like equations can be obtained from Eqs.(79), (80), (81) and (82) respectively.

The $\mathcal{D}$ equation of the quantum mechanics can be defined as
$$\mathcal{D} = (b\diamond)^* \circ (\mathcal{X} / b) \tag{88}$$

The $\mathcal{G}$ equation of the quantum mechanics can be defined as
$$\mathcal{G} = (b\diamond) \circ (\mathcal{D} / b) \tag{89}$$

The $\mathcal{T}$ equation of the quantum mechanics can be defined as
$$\mathcal{T} = (b\diamond)^* \circ (\mathcal{G} / b) \tag{90}$$

The $\mathcal{O}$ equation of the quantum mechanics can be defined as
$$\mathcal{O} = (b\diamond) \circ (\mathcal{T} / b) \tag{91}$$

Table 12.   Quantum equations set of trigintaduonion field T-C

| | |
|---|---|
| Energy quantum | $\mathcal{U} = (b\diamond)^* \circ (\mathcal{M} / b)$ |
| Power quantum | $\mathcal{L} = (b\diamond) \circ (\mathcal{U} / b)$ |
| Field potential quantum | $\mathcal{D} = (b\diamond)^* \circ (\mathcal{X} / b)$ |
| Field strength quantum | $\mathcal{G} = (b\diamond) \circ (\mathcal{D} / b)$ |
| Field source quantum | $\mathcal{T} = (b\diamond)^* \circ (\mathcal{G} / b)$ |
| Force quantum | $\mathcal{O} = (b\diamond) \circ (\mathcal{T} / b)$ |

## 8. Conclusions

By analogy with the four sorts of octonionic fields and twelve sorts of sedenion fields, four sorts of trigintaduonion fields and their special case have been developed, including their field equations, quantum equations and some new unknown particles.

In trigintaduonion field T-X, the study deduces the Dirac equation, Schrodinger equation, Klein-Gordon equation and some newfound equations of sub-quarks etc. It infers four sorts of Dirac-like equations of intermediate particles among sub-quarks etc. It predicts that there are some new particles of field sources (sub-quarks etc.) and their intermediate particles.

In trigintaduonion field T-A, the paper draws the Yang-Mills equation, Dirac equation, Schrodinger equation and Klein-Gordon equation of the quarks and leptons etc. It infers three sorts of Dirac-like equations of intermediate particles among quarks and leptons. It draws some conclusions of field source particles and intermediate particles which are consistent with current electro-weak theory. It predicts that there are some new unknown particles of field sources (quarks and leptons) and their intermediate particles.

In trigintaduonion field T-B, the research infers the Dirac equation, Schrodinger equation, Klein-Gordon equation and some newfound equations of electrons and masses etc. It deduces two sorts of Dirac-like equations of intermediate particles among electrons and masses etc. It draws some conclusions of field source particles and intermediate particles which are consistent with current electromagnetic and gravitational theories etc. It predicts that there are some new unknown particles of field sources (electrons and masses etc.) and their intermediate particles.

In trigintaduonion field T-S, the thesis concludes the Dirac equation, Schrodinger equation and Klein-Gordon equation of the galaxies etc. It infers Dirac-like equation of intermediate particles among galaxies. It predicts that there are some new unknown particles of field sources and their intermediate particles.

In the trigintaduonion field theory, we can find that the interplays among all eight sorts of interactions are much more mysterious and complicated than we found and imagined before.

**Acknowledgements**

The author thanks Shaohan Lin, Minfeng Wang, Yun Zhu, Zhimin Chen and Xu Chen for helpful discussions. This project was supported by National Natural Science Foundation of China under grant number 60677039, Science & Technology Department of Fujian Province of China under grant number 2005HZ1020 and 2006H0092, and Xiamen Science & Technology Bureau of China under grant number 3502Z20055011.